\documentclass[twocolumn,prd,amsmath,amssymb,preprintnumbers,reprint,showpacs]{revtex4}
\usepackage{epsfig}
\usepackage{textcomp}
\usepackage{subfigure}

\def\beq{\begin{equation}}
\def\eeq{\end{equation}}
\def\bea{\begin{eqnarray}} 
\def\eea{\end{eqnarray}}
\def\nn{\nonumber}

\def\mev{{\rm MeV}}
\def\kev{\rm keV}

\def\eps{\varepsilon}

\def\br{{\tt Br}}

\newcommand{\lsim}{
\mathrel{\hbox{\rlap{\hbox{\lower4pt\hbox{$\sim$}}}\hbox{$<$}}}}
\newcommand{\gsim}{
\mathrel{\hbox{\rlap{\hbox{\lower4pt\hbox{$\sim$}}}\hbox{$>$}}}}

\begin{document}

\preprint{CERN-PH-TH-2014-157}
\title{\boldmath Muon $g-2$ Anomaly and Dark Leptonic Gauge Boson}
\author{Hye-Sung Lee}
\affiliation{
Department of Physics, College of William and Mary, Williamsburg, Virginia 23187, USA \\
Theory Center, Jefferson Lab, Newport News, Virginia 23606, USA \\
Theory Division, CERN, CH-1211 Geneva 23, Switzerland
}
\date{August 2014}
\begin{abstract}
One of the major motivations to search for a dark gauge boson of MeV - GeV scale is the long-standing $g_\mu - 2$ anomaly.
Because of active searches such as fixed target experiments and rare meson decays, the $g_\mu - 2$ favored parameter region has been rapidly reduced.
With the most recent data, it is practically excluded now in the popular dark photon model.
We overview the issue and investigate a potentially alternative model based on the gauged lepton number or $U(1)_L$.
The $g_\mu - 2$ favored parameter region of the $U(1)_L$ survives all the constraints that were critical in the dark photon case, yet it is disfavored by the new constraints from the large flux neutrino experiments.
\end{abstract}
\maketitle

\section{Introduction}
The long-standing issue of the $g_\mu - 2$ has always been a great motivation and the constraint on the new physics beyond the Standard Model (SM).
It is one of the major motivations for the MeV - GeV scale gauge boson ($Z'$).
Through a loop correction, it can address the $3.6 \sigma$ C.L. discrepancy in the $g_\mu - 2$ \cite{Beringer:1900zz}.
There are also other motivations for such a light gauge boson in the context of the dark matter (DM) physics \cite{ArkaniHamed:2008qn} including the observation of the positron excess in the cosmic ray experiments \cite{Adriani:2008zr,Aguilar:2013qda}.
For a recent overview of the light dark gauge boson physics, see Ref.~\cite{Essig:2013lka}.
Unlike most other motivations, the $g_\mu - 2$ anomaly is independent of the unknown DM properties, and it is also independent of the unknown decay branching ratio of the $Z'$.

The popular approach to realize this light gauge boson with small couplings is often called the dark photon model \cite{ArkaniHamed:2008qn}, which exploits the small kinetic mixing of the new gauge symmetry $U(1)_\text{Dark}$ and the $U(1)_Y$ of the SM, $(\eps / 2 \cos\theta_W) \, B^{\mu\nu} Z'_{\mu\nu}$ \cite{Holdom:1985ag}.
In this model, the new gauge symmetry $U(1)_\text{Dark}$ is viewed as a gauge symmetry for the dark sector only.
The $Z'$ can still couple to the SM particles through the mixing with the SM gauge bosons.
Many searches for the dark gauge boson \cite{Essig:2013lka} have been motivated and analyzed in the context of this model \footnote{We note an extra $Z$-$Z'$ mass mixing may alter the phenomenological implications of the $Z'$ significantly \cite{Davoudiasl:2012ag,Davoudiasl:2012qa,
Lee:2013fda,Davoudiasl:2014mqa,Kong:2014jwa,Davoudiasl:2014kua}.}.

It is quite obvious that another way to naturally realize it is through a gauged lepton number ($L$) or $U(1)_L$ gauge symmetry as the $g_\mu - 2$ anomaly (in fact, the aforementioned positron excess as well) requires the $Z'$ to couple to the leptons, but not necessarily to the quarks.
The relevant experimental constraints and their interpretations differ from the kinetic mixing case (dark photon).

In fact, the entire $g_\mu - 2$ favored parameter space in the dark photon model is practically excluded by this year's experimental results as demonstrated in this paper.
Therefore, it would be timely and exigent to consider the $U(1)_L$ and examine the experimental constraints on the $g_\mu - 2$ favored parameter space of the model, which is the main purpose of this paper.
As we will show, it survives the constraints that excluded the dark photon as a solution to the $g_\mu - 2$ anomaly, but it does not survive the new constraints from the $\nu - e$ scattering experiments.

\begin{figure*}[tb]
\begin{center}
\subfigure[]{
      \includegraphics[width=0.47\textwidth,clip]{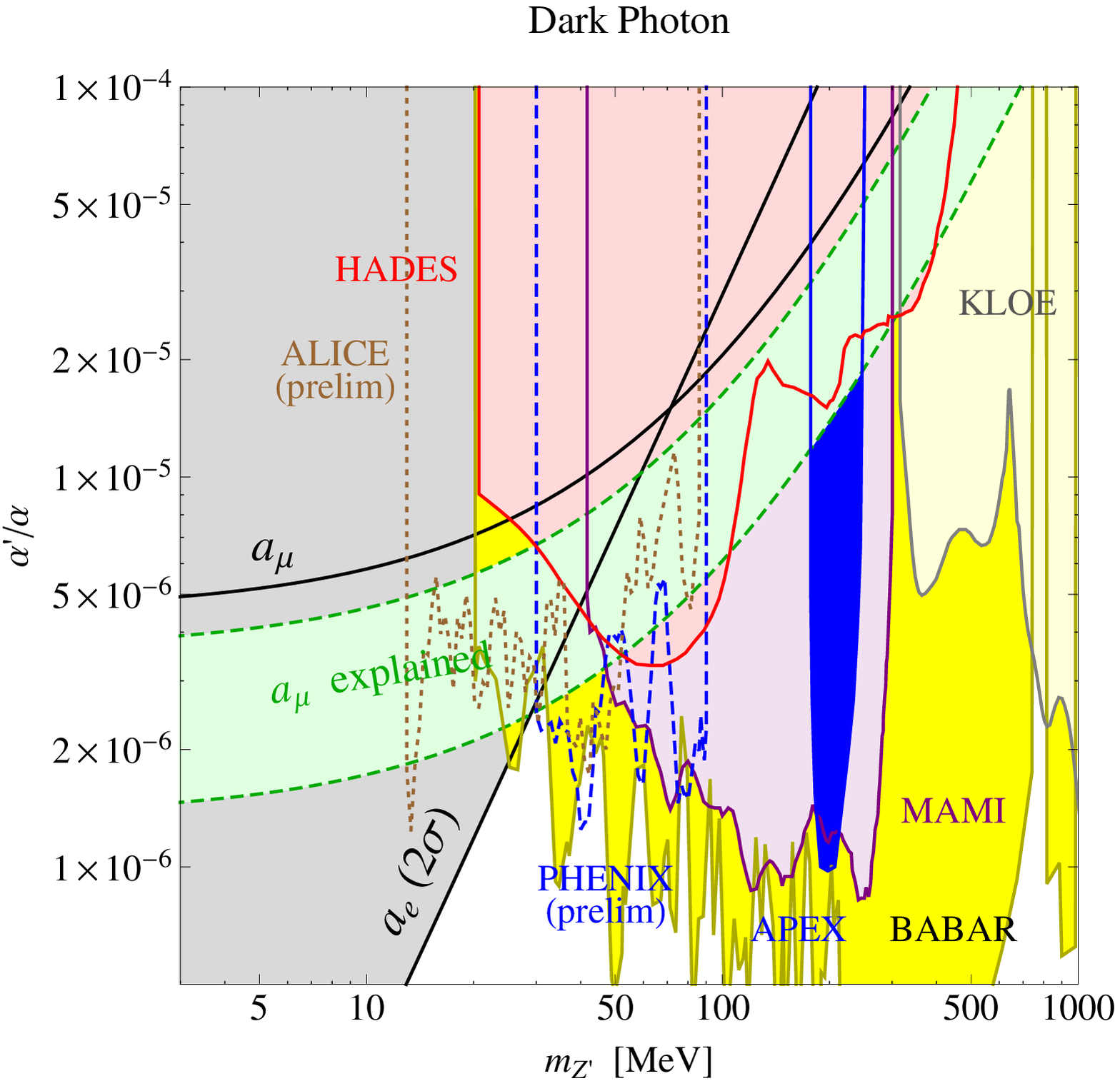}
} ~
\subfigure[]{
      \includegraphics[width=0.47\textwidth,clip]{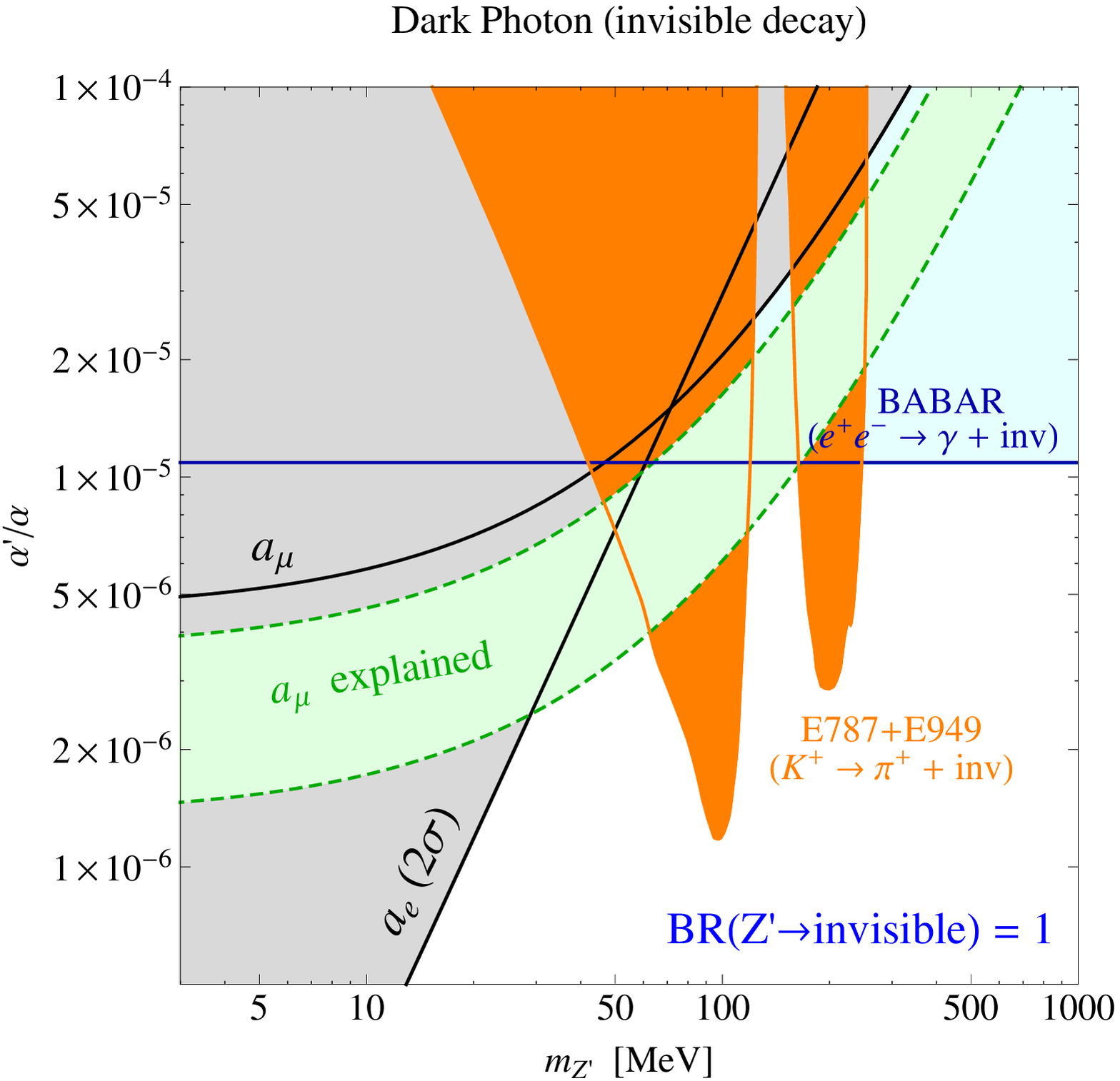}
}
\caption{Constraints on the dark photon parameter space (a) for the typical case of $\br(Z' \to \ell^+\ell^-) \approx 1$, and (b) for the very light dark matter case of $\br(Z' \to \text{invisible}) = 1$.
The figures are based on Ref.~\cite{Davoudiasl:2014kua} with the updates from new data and cosmetic changes.
The $g_\mu - 2$ favored region (green band) in typical case (a) is practically closed with the new data.
}
\label{fig:parameterspace}
\end{center}
\end{figure*}

\section{Dark Leptonic Gauge Boson}
Light leptonic gauge bosons, including the flavor-dependent ones, have been studied in numerous articles \cite{Baek:2001kca,Joshipura:2003jh,Heeck:2010pg,Williams:2011qb,Batell:2011qq,Barger:2011mt,Feng:2012jn,Carone:2013uh,Agrawal:2014ufa,Fayet}.
In this paper, we consider the gauged $U(1)_L$ with a gauge boson $Z'$ of the GeV or sub-GeV scale.
We show an example of the anomaly-free particle spectrum in the Appendix.
Some details regarding the new particles are also presented there although they do not affect the main phenomenology we consider in this paper.
While we do not necessarily link our discussion to the dark matter physics, we call our $Z'$ the {\em dark leptonic gauge boson} as the word ``dark'' is occasionally used for the suppressed coupling in contrast to the ``bright'' photon coupling.

The $Z'$ interactions with the SM fermions in each model, at the leading order, are given by
\bea
{\cal L}_\text{dark photon} &=& - \eps e Q_e (f) \, \bar f \gamma^\mu f \, Z'_\mu \, , \\
{\cal L}_\text{dark leptonic gauge boson} &=& - g_{Z'} Q_L (f) \, \bar f \gamma^\mu f \, Z'_\mu \, ,
\eea
where $Q_e$ is the electric charge and $Q_L$ is the lepton number.
In the $U(1)_L$ model with a small gauge coupling, the light mass of the $Z'$ arises naturally from the small gauge coupling by $m_{Z'} \sim g_{Z'} \left< S \right>$ or similar depending on the symmetry breaking sector.

Although the $g_{Z'}$ may have a very small size compared to the SM gauge couplings, such a small gauge coupling constant is technically fine as it neither violates any symmetry nor generates a fine-tuning issue.
In fact, the difference may not be particularly drastic in view of the disparity between the Yukawa coupling sizes of the top quark and the electron in the SM (about $10^6$ times difference).
The precision test of the $Z$ pole location at the LEP experiment, which is constrained to be less than ${\cal O} (10^{-3})$ in terms of the $Z$-$Z'$ mixing angle \cite{Beringer:1900zz}, is easily satisfied with the small $g_{Z'}$.

The relevant $Z'$ decay channels are
\bea
m_{Z'} < 2 m_e &:& \nu_i \bar \nu_i \\
2 m_e < m_{Z'} < 2 m_{\mu} &:& \nu_i \bar \nu_i ,\, e^+ e^- \\
2 m_\mu < m_{Z'} < 2 m_{\tau} &:& \nu_i \bar \nu_i ,\, e^+ e^- ,\, \mu^+ \mu^-
\eea
with $i = e$, $\mu$, $\tau$.
For a large part of the $m_{Z'}$ range, the phase space suppression can be neglected.
For example, the decay branching ratios into the neutrinos are about 75\% in $1 ~\mev \lsim m_{Z'} \lsim 200 ~\mev$.
The leptonic gauge boson is potentially a great source of the neutrino production in all flavors.

\section{Anomalous Magnetic Moment}
There is a $3.6 \sigma$ discrepancy in the muon anomalous magnetic moment $a_\mu \equiv (g_\mu - 2) / 2$ between the experimental value and the SM prediction \cite{Bennett:2006fi,Beringer:1900zz}.
\beq
\Delta a_\mu = a_\mu^\text{exp} - a_\mu^\text{SM} = 288 (80) \times 10^{-11}
\eeq
This can be explained by the $Z'$ with the dominant vector coupling to leptons \cite{Gninenko:2001hx,Fayet:2007ua,Pospelov:2008zw}.
The contribution to $a_\ell$, for the lepton $\ell$, by the dark gauge boson is
\bea
&& a_\ell^{Z'} = \frac{\alpha'}{2\pi} F_V \left( m_{Z'} / m_\ell \right) \quad \text{with} \\
&& F_V(x) \equiv \int_0^1 dz [2 z (1-z)^2] / [(1-z)^2 + x^2 z] \, ,
\eea
where $\alpha' \equiv \eps^2 \alpha$ (dark photon) or $\alpha' \equiv g_{Z'}^2 / 4\pi$ (dark leptonic gauge boson).

In Fig.~\ref{fig:parameterspace}, we plot the parameter region that can explain the $\Delta a_\mu$ in 90\% C.L. (green band) as well as $3 \sigma$ C.L. exclusion bound.
We also plot the bound from $a_e$ in $2 \sigma$ C.L. \cite{Davoudiasl:2012ig,Endo:2012hp} using the up-to-date value.
\beq
\Delta a_e = a_e^\text{exp} - a_e^\text{SM} = -1.06 (0.82) \times 10^{-12}
\eeq

\section{Constraints on the Dark Photon}
Now, we overview the current bounds in the popular dark photon model.
Typically, there are two kinds of the dark photon searches in the labs:

$~~$ (i) Dilepton resonance search ($Z' \to \ell^+ \ell^-$),

$~~$ (ii) Missing energy search ($Z' \to \chi \bar \chi$).

The dilepton resonance search is the most popular search in the ordinary condition, while the missing energy search assumes the existence of the light dark matter (LDM) whose mass is $m_{\chi} < m_{Z'} / 2$ so that the dark photon can dominantly decay into them.
Since the dark photon couples only to the electrically charged particles, it does not couple to the SM neutrinos that would have appeared as a missing energy.
Different types of experimental constraints apply in each case.

Figure~\ref{fig:parameterspace}~(a) shows the parameter space of the dark photon model in the ordinary case, i.e., without assuming the existence of the LDM particles.
The dilepton decay branching ratio is large in this model over most of the mass range \cite{Batell:2009yf,Falkowski:2010cm}.
For $m_{Z'} < 2 m_\mu$, it has $\br (Z' \to e^+ e^-) \simeq 1$.
Various constraints from the meson decays \cite{Bjorken:2009mm,Agakishiev:2013fwl,Babusci:2014sta,Lees:2014xha,PHENIX,ALICE}, fixed target experiments \cite{Abrahamyan:2011gv,Merkel:2014avp}, beam dump experiments \cite{Andreas:2012mt} apply, and only the relevant ones for the $g_\mu - 2$ favored region (green band) are shown in the figure.
We refer the interested readers to Ref.~\cite{Essig:2013lka} for the wider parameter space and relevant bounds.

With this year's results from MAMI (90\% C.L.) \cite{Merkel:2014avp} and BABAR (90\% C.L.) \cite{Lees:2014xha} along with the $a_e$ bound ($2 \sigma$ C.L.), we can see the whole green band is practically excluded except for a tiny spot, which is also closed when the preliminary results from PHENIX (90\% C.L.) \cite{PHENIX} and ALICE (90\% C.L.) \cite{ALICE} are counted.
Thus one of the major motivations of the dark gauge boson is seriously weakened now in the popular dark photon model.

Partly because of the issue of the rapidly shrinking green band, there has been a growing interest in the invisibly-decaying dark photon with the LDM particles.
Figure~\ref{fig:parameterspace}~(b) shows the parameter space of the dark photon model with $\br(Z' \to \chi \bar \chi) = 1$.
Here, the $K^+ \to \pi^+$ + nothing search (95\% C.L.) based on the BNL E787+E949 results \cite{Artamonov:2009sz} and the off-resonance $e^+e^- \to \gamma$ + nothing search (95\% C.L.) \cite{Izaguirre:2013uxa,Essig:2013vha} based on the BABAR results \cite{Aubert:2008as} provide strong constraints.

Although sizable parameter space is excluded, there are still some portions of the green band around $m_{Z'} \approx 30 - 50 ~\mev$ and $m_{Z'} \approx 140 ~\mev$ that survive the constraints.
There may be more constraints using the LDM detection such as the E137 beam dump experiments \cite{Batell:2014mga}, but they depend on other parameters such as the LDM coupling, and we do not consider them in this paper.
Part of the unconstrained parameters in this scenario can be probed by the proposed invisibly-decaying dark gauge boson searches at DarkLight \cite{Kahn:2012br} and SPS \cite{Gninenko:2013rka,Andreas:2013lya}.

\section{Physics of the Dark Leptonic Gauge Boson}
A drawback of the invisibly-decaying dark photon [Fig.~\ref{fig:parameterspace} (b)] is the requirement of the very light dark matter sector particles, lighter than the MeV - GeV scale dark photon.
The dark leptonic gauge boson has the SM neutrinos the $Z'$ can mainly decay into, and it is free from the severe quarkonium decay constraints \cite{Agakishiev:2013fwl,Babusci:2014sta,Lees:2014xha,PHENIX,ALICE} as it does not couple to quarks.
It is worthwhile investigating the constraints on its green band.

The relevant experiments for the dark leptonic gauge boson include (i) Fixed target experiments, (ii) BABAR $e^+e^- \to \gamma$ + nothing search, (iii) BABAR $e^+e^- \to \gamma + \ell^+\ell^-$, (iv) Neutrino trident experiments, (v) SLAC E137 experiment (LDM search), and (vi) Beam dump experiments.
Most of these were investigated to constrain the dark photon (Fig.~\ref{fig:parameterspace}) or other scenarios, and we can use them with appropriate interpretations for the dark leptonic gauge boson.

\begin{figure}[tb]
\begin{center}
\includegraphics[width=0.47\textwidth]{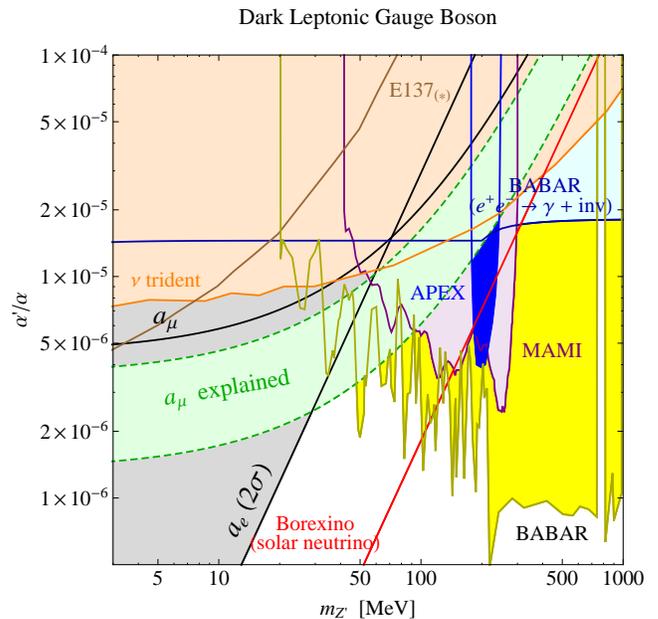}
\caption{Constraints on the parameter space of the dark leptonic gauge boson in the $U(1)_L$ model.
}
\label{fig:U1Lparameterspace}
\end{center}
\end{figure}

(i) Fixed target experiments (APEX \cite{Abrahamyan:2011gv}, MAMI \cite{Merkel:2014avp} with 90\% C.L.):
They look for the $e^+ e^-$ resonance from the bremsstrahlung ($e N \to e N + e^+e^-$).
Since the $Z'$ decay branching ratio for the $e^+ e^-$ is diluted due to the SM neutrinos, the results in Fig.~\ref{fig:parameterspace}~(a) can be taken for the dark leptonic gauge boson with a proper scaling.
The scaling is about $1/4$ for $m_{Z'} < 2 m_\mu$.

(ii) BABAR off-resonance $e^+e^- \to \gamma$ + nothing search:
They look for the $\gamma$ + missing energy as a signal of the invisibly decaying $Z'$ into the LDM particles in the dark photon scenario (95\% C.L.) \cite{Izaguirre:2013uxa,Essig:2013vha}.
In the dark leptonic gauge boson case, the SM neutrinos play the same role as the LDM for the missing energy.
The results in Fig.~\ref{fig:parameterspace} (b) can be taken with a proper scaling.
Unlike the dark photon, the dark leptonic gauge boson couples to the SM neutrinos and the charged leptons with the same strength.
The scaling is about $3/4$ for $m_{Z'} < 2 m_\mu$.

(iii) BABAR $e^+e^- \to \gamma + \ell^+\ell^-$ search:
BABAR recently released a new analysis of the $\gamma + \ell^+\ell^-$ ($\ell = e$, $\mu$) channel for the dark photon \cite{Lees:2014xha}.
Unlike most other meson analyses including the old BABAR constraint \cite{Bjorken:2009mm}, the new BABAR result assumes the dark photon production only from the leptons, and not through the meson decays based on the idea that a vector meson decay to a photon and a light vector boson should be suppressed.
Thus the new BABAR constraint (90\% C.L.)  does not vanish for the dark leptonic gauge boson.
The scaling is about $1/4$ for $m_{Z'} < 2 m_\mu$.

(iv) Neutrino trident ($\nu_\mu N \to \nu_\mu N + \mu^+ \mu^-$) experiments:
The bound (95\% C.L.) from the neutrino trident production based on the CCFR experiment \cite{Mishra:1991bv} is given for the $L_\mu - L_\tau$ model in Ref.~\cite{Altmannshofer:2014pba}.
They look for the $\mu^+ \mu^-$ pair produced in the muon neutrino beam scattered at the target.
It provides a strong constraint on the $Z'$ which couples to both muons and muon neutrinos.
Their bounds can be directly used for the $U(1)_L$ based model without scaling.

(v) SLAC E137 experiment (LDM search):
The bound for the LDM (95\% C.L.) from the E137 beam dump experiment \cite{Bjorken:1988as} can be found in Ref.~\cite{Batell:2014mga}.
They analyzed the data for the dark photon decaying into the fermionic LDM (with $m_\chi < 0.5 ~\mev$ and LDM coupling $\alpha_D = 0.1$), which may scatter with the electrons in the detector via the dark photon.
We can reinterpret their LDM bounds for the SM neutrinos with an appropriate scaling.
While $\br(Z' \to \chi \bar\chi) = 1$ is presumed for the LDM, the branching ratio for the SM neutrinos in the $U(1)_L$ is smaller (about $3/4$ for $m_{Z'} < 2 m_\mu$).
The expected number of signals scales as $(\eps^2 \alpha)^2 \alpha_D$ for the LDM, while it scales as $\alpha'^3$ for the neutrinos.
Unlike the dark photon with the LDM, which has $\alpha_D$ dependence, the $U(1)_L$ does not have an extra parameter.
There are also $Z / W$ mediated scatterings, which we neglect here as they are subdominant for the parameter region we are interested in.

(vi) Other beam dumps:
There are numerous beam dump experiments in various labs looking for the long-lived particles decaying into $e^+e^-$ eventually \cite{Andreas:2012mt}.
This kind of experiment is only sensitive to the $Z'$ that is long-lived enough to pass through the thick shields, which is the main reason why they are not very relevant in constraining the green band.
Because of the SM neutrinos, the decay width of the dark leptonic gauge boson is larger than that of the dark photon (about 4 times for $m_{Z'} < 2 m_\mu$).
Thus, its decay branching ratio into the $e^+ e^-$ pair as well as the displacement vertex is about $1/4$ of that of the dark photon, for $m_{Z'} < 2 m_\mu$.
The chance to pass through the shield and get detected is greatly suppressed for the dark leptonic gauge boson, and the beam dump bounds are much weaker than the dark photon case, which are already irrelevant to the green band.

The dark leptonic gauge boson can also mediate the scattering of neutrinos and electrons, which would be important especially in the low-energy experiments.
In fact, the large flux neutrino experiments are quite sensitive to such a $\nu - e$ scattering \cite{Harnik:2012ni,Laha:2013xua}.
The bounds from the $862 ~\kev$ $^7$Be solar neutrino flux measurement at the Borexino experiment \cite{Bellini:2011rx} for the $B-L$ model are given in Ref.~\cite{Harnik:2012ni}, which already excludes the $g_\mu - 2$ favored parameter region.
This can also be taken for the $U(1)_L$.

In Fig.~\ref{fig:U1Lparameterspace}, we show the parameter space of the dark leptonic gauge boson with the constraints discussed above.
Although more dedicated studies, instead of using the simple scalings, might reveal some differences, Fig.~\ref{fig:U1Lparameterspace} has sufficient details and accuracy for our purpose.
The dark leptonic gauge boson of roughly $m_{Z'} \approx 30 - 80 ~\mev$ range (which is the parameter range sensitive to the proposed fixed target experiments using the electron beams at JLab \cite{McKeown:2011yj,Kahn:2012br} and Mainz \cite{Merkel:2014avp}) survives all constraints, but not the constraint from the $\nu - e$ scattering.
This scattering is irrelevant for the dark photon which couples only to the electromagnetic current, but is very important for the dark leptonic gauge boson.
Although it might still be meaningful to perform other types of experiments, the solar neutrino experiment results give the discouragingly strong bounds for the $U(1)_L$ based dark leptonic gauge boson as a solution to the $g_\mu - 2$ anomaly.

Regardless of being the solution to the $g_\mu - 2$ anomaly, however, the dark leptonic gauge boson predicts interesting physics worth investigating.
Because of the large branching ratio into the SM neutrinos (100\% if $m_{Z'} < 1 ~\mev$), the dark leptonic gauge boson produced via bremsstrahlung from the electron (such as in the fixed target or beam dump experiments) would effectively serve as a neutrino beam producing machine, which deserves a dedicated study \cite{Lee}.
This is a feature for any light gauge boson that couples to both electrons and neutrinos in general, and it does not apply to the $L_\mu - L_\tau$ model \cite{Altmannshofer:2014pba}.
The neutrino beam from the $U(1)_L$ would have distinguishable properties such as the presence of both $\nu$ and $\bar \nu$ of all three flavors in equal portion.

\section{Summary and Conclusions}
The $g_\mu - 2$ anomaly at the $3.6\sigma$ C.L. provides a great motivation for the light dark gauge boson, independent of the unknown dark matter properties.
However, as we demonstrated in this paper, the popular dark photon model is seriously hampered as the $g_\mu - 2$ favored parameter region is excluded by the recent experimental results with the described confidence levels unless very light dark matter particles are introduced.

We investigated the $U(1)_L$ based dark leptonic gauge boson as a potential alternative to the dark photon, as it contains all essential features to explain the $g_\mu - 2$ in the minimal flavor-universal way.
It preserves the $g_\mu - 2$ favored parameter region after all the experimental constraints that are sensitive to the dark photon, but it does not survive the $\nu - e$ scattering bounds set by the Borexino $^7$Be solar neutrino flux measurement.
Thus, it would be hard to consider the $U(1)_L$ as a solution to the $g_\mu - 2$ anomaly although it would still be worthwhile to study the $U(1)_L$ based dark gauge boson.

Further aspects to study about the $U(1)_L$ based dark leptonic gauge boson include the phenomenology of the exotic leptons, required for the gauged $U(1)_L$, especially for the Large Hadron Collider experiments and detailed implications for the neutrino physics.

\begin{acknowledgments}
We appreciate hospitality from Seoul Tech, Yonsei University, and KIAS during the visits.
We thank C. Carone for the useful discussion about anomaly-free conditions, J. Heeck for the useful discussion about the neutrino scattering bounds, and H. Davoudiasl and W. Marciano for the extensive discussions about the light gauge boson physics.
We also thank B. Echenard for the useful correspondence about the recent BABAR analysis.
This work was supported in part by the U.S. DOE under Grant No.~DE-AC05-06OR23177, the NSF under Grant No.~PHY-1068008, and the CERN-Korea fellowship.
\end{acknowledgments}

\appendix
\section{\boldmath Anomaly-free $U(1)_L$ example}
Here, we present an example of the anomaly-free particle spectrum for the gauged $U(1)_L$.
(See also Refs.~\cite{FileviezPerez:2010gw,Chao:2010mp,Dong:2010fw,Ko:2010at,Duerr:2013dza,Schwaller:2013hqa}.)
The particle contents of the model is the whole group of SM particles including three right-handed neutrinos, two sets of entire exotic lepton multiplet $(L_4, N_4, E_4)$, $(L_5, N_5, E_5)$ which are vector-like under the SM gauge group, and the  $Z'$ as well as a Higgs singlet $S$ to break the $U(1)_L$.
We use notations such as $L \equiv (\nu, e)_L^T$, $N \equiv \nu_R$, $E \equiv e_R$.
The charge assignments are given in Table~\ref{tab:charges}.
The SM neutrinos are Dirac in this model as the Majorana neutrinos would have vanishing vector couplings to the vector boson.

The SM leptons (quarks) have $1$ ($0$) for the new $U(1)$ charges, justifying the name $U(1)_L$ for the SM sector.
The exotic-lepton mediated mixing of the $Z'$ with the SM neutral gauge bosons would be highly suppressed because of the small $g_{Z'}$.
The model is free from all chiral anomalies as well as the global anomaly condition that requires an even number of doublet fermions.

\begin{table}[t]
\begin{tabular}{|cc||ccc|c|}
\hline
& ~Field~ & $SU(3)_C$ & $SU(2)_L$ & $U(1)_Y$ & $U(1)_L$  \\
\hline\hline
                  & $Q_{1 - 3}$ & $3$ & $2$ & $~~\frac{1}{6}$  & $0$ \\
SM quarks & $U_{1 - 3}$ & $3$ & $1$ & $~~\frac{2}{3}$  & $0$ \\
                  & $D_{1 - 3}$ & $3$ & $1$ & $-\frac{1}{3}$ & $0$ \\
\hline
                   & $L_{1 - 3}$ & $1$ & $2$ & $-\frac{1}{2}$ & $1$ \\
SM leptons & $N_{1 - 3}$ & $1$ & $1$ & $~~0$  & $1$ \\
                   & $E_{1 - 3}$ & $1$ & $1$ & $-1$ & $1$ \\
\hline
                       & $L_{4}$ ($L_{5}$)    & $1$ & $2$ & $-\frac{1}{2}$ ($\frac{1}{2}$) & $-3$ ($0$) \\
Exotic leptons & $N_{4}$ ($N_{5}$)  & $1$ & $1$ & $~~0$ ($0$)  & $-3$ ($0$) \\
                       & $E_{4}$ ($E_{5}$)   & $1$ & $1$ & $-1$ ($1$) & $-3$ ($0$) \\
\hline
SM Higgs      & $H$      & $1$ & $2$ & $~~\frac{1}{2}$ & $0$ \\
Higgs singlet & $S$      & $1$ & $1$ & $~~0$               & $3$ \\
\hline
\end{tabular}
\caption{$SU(3)_C \times SU(2)_L \times U(1)_Y \times U(1)_L$ charges of the fields in the model. The exotic leptons carry unusual lepton numbers for the chiral anomaly cancellation.}
\label{tab:charges}
\end{table}

The Yukawa part Lagrangian is
\begin{align}
~&-{\cal L}_\text{Y} = y_U \bar Q \tilde H U + y_D \bar Q H D + y_L \bar L \tilde H N + y_E \bar L H E \nn \\
&+ y_{N_4} \bar L_4 \tilde H N_4 + y_{N_5} \bar L_5 H N_5 + y_{E_4} \bar L_4 H E_4  + y_{E_5} \bar L_5 \tilde H E_5 \nn \\
&+ c_L S L_4 L_5 + c_N S N_4 N_5 + c_E S E_4 E_5 + h.c.
\end{align}
where $\tilde H \equiv i \tau_2 H^*$ is a conjugate of $H$.
This is a very similar particle spectrum [except for the new $U(1)_L$ charges] to those in Refs.~\cite{ArkaniHamed:2012kq,Davoudiasl:2012ig}.
In Ref.~\cite{Davoudiasl:2012ig}, all new leptons have pure vector couplings to the gauge boson.
For instance, $L_4$ and $L_5$ there have equal opposite charges under the new $U(1)$ gauge symmetry.
The $U(1)_L$ charges of the $L_4$ and $L_5$ differ in our model.

There are two charged Dirac fermions $\ell'^\pm_1$, $\ell'^\pm_2$ of a unit electric charge $\pm 1$ and two neutral Dirac fermions $\nu'_1$, $\nu'_2$.
Although $N_5$ does not contribute to any anomaly condition as it is a singlet under both the SM gauge group and the $U(1)_L$, we add it so that the new charged particles can decay into $\nu'_1 + W$ \cite{ArkaniHamed:2012kq,Davoudiasl:2012ig}.
Otherwise there would be only one neutral Dirac particle that is heavier than the lightest charged one $\ell'_1$, which would be severely constrained.
For the discussion on whether the $\nu'_1$ can be the relic DM candidate, see Ref.~\cite{Davoudiasl:2012ig}.



\end{document}